\documentclass[conference]{IEEEtran}
\usepackage{amsmath,amssymb,multicol,graphicx,gensymb,flushend,booktabs,array,url}
\IEEEoverridecommandlockouts

\begin{document}

\title{Ceramic Materials: From Atomic Bonding and Processing to Microstructure, Functional Properties, and Biomedical Applications}

\author{\IEEEauthorblockN{Jairo Rond\'on\IEEEauthorrefmark{1},
Juan Valera\IEEEauthorrefmark{2}, Alex V\'elez Cruz\IEEEauthorrefmark{2}, \& Irela Perez Magin\IEEEauthorrefmark{2}}
\IEEEauthorblockA{\IEEEauthorrefmark{1}Biomedical \& Chemical Engineering Departments,\
Polytechnic University of Puerto Rico, San Juan, Puerto Rico, USA\\jrondon@pupr.edu}
\IEEEauthorblockA{\IEEEauthorrefmark{2}Biomedical Engineering Department,\
Polytechnic University of Puerto Rico, San Juan, Puerto Rico, USA\\
jvalera@pupr.edu, alvelez@pupr.edu, iperezmagin@pupr.edu}}

\maketitle

\begin{abstract}
Ceramic materials are inorganic, predominantly nonmetallic solids whose performance is governed by the interplay among atomic bonding, crystal structure, processing, microstructure, and functional properties. Their ionic and covalent bonding provides high hardness, thermal and chemical stability, wear resistance, and diverse electrical behavior, but also limits plastic deformation and increases sensitivity to flaws, porosity, grain boundaries, and residual stresses. This review examines ceramics through the processing--microstructure--property--performance relationship, emphasizing the effects of crystal chemistry, defects, grain size, porosity, and phase stability on mechanical, thermal, electrical, and biological behavior. Conventional sintering, sol--gel processing, and additive manufacturing are discussed as key routes for controlling densification, porosity, geometry, and composition. Biomedical ceramics, including alumina, zirconia, calcium phosphates, hydroxyapatite, tricalcium phosphate, and bioactive glasses, are also reviewed. Emerging topics include nanoceramics, smart ceramics, composites, multimaterial manufacturing, defect engineering, immunomodulatory bioceramics, and sustainable processing. Major challenges remain in brittle fracture, processing heterogeneity, degradation control, scale-up, and the porosity--strength trade-off. Future advances will require integrating advanced manufacturing, multiscale characterization, predictive modeling, and data-driven design.
\end{abstract}

\section{Introduction}
Ceramic materials have accompanied technological development for thousands of years, evolving from clay-based pottery and structural components into a highly diversified family of engineering materials used in energy systems, electronics, aerospace technologies, environmental applications, dentistry, orthopedics, regenerative medicine, and advanced manufacturing \cite{r1,r2,r3}. The contemporary concept of a ceramic extends far beyond traditional silicate-based materials and encompasses inorganic, predominantly nonmetallic solids whose properties are governed by combinations of ionic and covalent bonding, crystalline or amorphous organization, chemical composition, defect populations, and processing history.

The performance of ceramic materials cannot be understood from composition alone. Instead, it emerges from a hierarchical relationship linking atomic-scale bonding to crystal structure, microstructure, processing, and macroscopic response:
$$
\text{Bonding} \rightarrow \text{Crystal Structure}\rightarrow\text{Microstructure}\rightarrow\\
$$
$$
\text{Properties}\rightarrow\text{Performance}
$$
Strong ionic and covalent bonds generally produce high bond energies, high melting or decomposition temperatures, chemical stability, and elevated elastic stiffness. At the same time, these bonding configurations strongly restrict dislocation-mediated plastic deformation compared with metallic systems. Consequently, many dense ceramics exhibit high hardness and compressive strength but remain sensitive to tensile stresses and pre-existing flaws \cite{r1,r4}.

Within linear elastic fracture mechanics, flaw sensitivity may be represented as
\begin{equation}
K_I=Y\sigma\sqrt{\pi a},
\end{equation}
where $K_I$ is the mode-I stress intensity factor, $\sigma$ is applied tensile stress, $a$ is a characteristic flaw size, and $Y$ is a geometric factor. Unstable crack propagation occurs when
\begin{equation}
K_I\geq K_{IC},
\end{equation}
where $K_{IC}$ is the critical fracture toughness. Therefore,
\begin{equation}
\sigma_f\propto\frac{K_{IC}}{\sqrt{\pi a}},
\end{equation}
illustrating why processing-induced flaws strongly affect ceramic strength and reliability \cite{r4,r5}.

Processing occupies a central position in ceramic science. Powder synthesis, particle-size distribution, agglomeration, forming pressure, binder content, green density, debinding conditions, sintering temperature, dwell time, atmosphere, and cooling rate can each modify pore structure, grain size, phase distribution, and residual stress \cite{r4,r6}. Modern routes such as sol--gel processing and additive manufacturing further expand geometric and compositional design freedom while introducing additional variables related to rheology, curing, binder removal, and anisotropic shrinkage \cite{r7,r8,r9,r10}.

Biomedical ceramics provide a compelling example of this integrated design problem. Alumina and stabilized zirconia are used primarily for wear resistance, chemical stability, and mechanical reliability, whereas calcium phosphates and bioactive glasses are selected when controlled interaction with physiological fluids and tissues is required \cite{r11,r12,r13}. Hydroxyapatite and tricalcium phosphate exhibit chemical affinity with bone mineral but are mechanically limited as monolithic load-bearing systems. Zirconia, in contrast, can exploit stress-induced tetragonal-to-monoclinic transformation for transformation toughening, although the same metastability also contributes to low-temperature degradation under humid conditions \cite{r14,r15,r16}.

The objective of this review is to provide an integrated analysis of ceramic materials based on the relationship among atomic bonding, crystal structure, processing, microstructure, properties, and technological performance. Particular emphasis is placed on biomedical ceramics because they simultaneously require mechanical reliability, surface chemistry, controlled degradation, and biological interaction. Conventional and advanced processing methods are examined together with structural, functional, and emerging ceramic systems, while current limitations and future directions involving defect engineering, nanostructuring, smart ceramics, multimaterial manufacturing, data-driven design, and sustainable processing are critically discussed.

\section{Methodology}
This review was developed using a structured literature-review and critical-analysis approach designed to examine ceramic materials from an integrated processing--microstructure--property--performance perspective. The literature basis was centered on peer-reviewed journal articles, review papers, academic books, and selected technical sources addressing structure, processing, properties, and applications of ceramic materials.

Search concepts were organized around combinations of the following terms: ceramic materials, advanced ceramics, ceramic microstructure, ceramic processing, sintering, sol--gel processing, ceramic additive manufacturing, bioceramics, hydroxyapatite, calcium phosphate ceramics, zirconia, alumina, bioactive glasses, nanoceramics, smart ceramics, ceramic composites, sustainable ceramics, and recycled ceramic feedstocks. The core time window emphasized literature from 2014--2026 while retaining foundational sources required to explain ceramic processing, fracture mechanics, and bioceramic mechanisms.

References were considered suitable when they directly supported at least one of the following areas: atomic bonding and crystal chemistry; phase stability and defects; microstructure--property relationships; mechanical, thermal, dielectric, or chemical behavior; conventional or advanced processing routes; additive manufacturing; biomedical ceramics; nanostructured or smart systems; ceramic composites; or sustainability and recycling. Priority was given to peer-reviewed work with clear experimental, theoretical, or review-based relevance.

The selected literature was organized according to
$$
\text{Composition and Bonding} \rightarrow \text{Processing} \rightarrow
$$
$$
\text{Microstructure} \rightarrow \text{Properties} \rightarrow \text{Performance}
$$
Each material system or process was examined at three levels: (i) identification of the underlying physicochemical mechanism, (ii) evaluation of how processing and microstructure modify that mechanism, and (iii) interpretation of the resulting technological or biomedical performance. This approach was selected to move beyond descriptive statements and to emphasize causal relationships between structure and behavior.

Processing methods were compared according to their ability to control densification, grain growth, residual porosity, pore connectivity, compositional homogeneity, dimensional accuracy, internal defects, geometric complexity, scalability, and suitability for biomedical customization. Biomedical systems were analyzed through a combined materials-science and biological-response framework:
$$
\text{Composition} \rightarrow \text{Phase Structure} \rightarrow \text{Surface}
$$
$$
\text{Chemistry} \rightarrow \text{Degradation/Stability} \rightarrow \text{Biological Interaction}
$$

The review does not claim to be a formal meta-analysis; rather, it provides a reproducible conceptual framework for comparing diverse ceramic systems whose testing conditions, compositions, and processing histories frequently differ substantially.

\section{Fundamentals of Ceramic Materials}
\subsection{Atomic Bonding and Crystal Structure}
The distinctive properties of ceramic materials originate fundamentally from their atomic bonding and crystallographic organization. Bonding in ceramic materials is predominantly ionic, covalent, or mixed ionic--covalent. In ionic ceramics, charge neutrality imposes compositional constraints that may be written as
\begin{equation}
\sum_i z_iN_i=0,
\end{equation}
where $z_i$ is ionic charge and $N_i$ is the corresponding population. In alumina, for example,
\begin{equation}
2(+3)+3(-2)=0.
\end{equation}
Such charge-balance requirements influence site occupancy, permissible substitutions, and defect chemistry \cite{r1,r2}.

Strongly covalent ceramics such as SiC and Si$_3$N$_4$ exhibit directional bonding, high bond energies, low atomic mobility, and excellent high-temperature stability. However, the same directional character limits the number of favorable slip systems and suppresses extensive room-temperature plasticity. Ceramic bonding is best viewed as a continuum rather than a strict ionic/covalent dichotomy \cite{r1,r2}.

Crystal structure represents the next hierarchical level controlling behavior. Ionic size, charge balance, coordination, electronic configuration, covalency, temperature, pressure, and defect chemistry collectively determine stable crystal arrangements. Perovskite is an important but non-universal ceramic structure, commonly represented by $ABO_3$. Other major structures include rock-salt, fluorite, spinel, corundum, wurtzite, zinc-blende, silicate networks, calcium-phosphate structures, and glassy networks \cite{r2,r17}.

Zirconia provides a key example of phase engineering. Pure ZrO$_2$ exhibits monoclinic, tetragonal, and cubic polymorphs, while additions of Y$_2$O$_3$, CeO$_2$, CaO, or MgO can stabilize tetragonal or cubic phases at lower temperatures. Stress-induced transformation
\begin{equation}
t\text{-ZrO}_2\rightarrow m\text{-ZrO}_2
\end{equation}
produces local volumetric expansion and compressive stresses around crack tips, thereby increasing fracture resistance \cite{r14,r15,r18}.

Hydroxyapatite, Ca$_{10}$(PO$_4$)$_6$(OH)$_2$, offers another example in which lattice chemistry influences function. Ionic substitutions at calcium, phosphate, or hydroxyl sites can modify lattice parameters, crystallinity, dissolution, surface chemistry, and biological response \cite{r12,r19}.

\subsection{Defects, Grain Boundaries, and Microstructure}
Real ceramics contain point defects, dislocations, grain boundaries, pores, secondary phases, inclusions, and microcracks. In ionic solids, point defects are constrained by electroneutrality. Oxygen-vacancy formation may be represented using Kr\"oger--Vink notation as
\begin{equation}
O_O^{\times}\rightarrow V_O^{\bullet\bullet}+\frac{1}{2}O_2(g)+2e'.
\end{equation}
The concentration and mobility of such defects affect ionic conductivity, redox behavior, dielectric loss, catalytic activity, and surface chemistry \cite{r1,r20}.

Grain boundaries are structurally and chemically distinct from grain interiors and strongly influence diffusion, segregation, fracture, electrical transport, sintering, and grain growth. In functional ceramics, local space-charge regions and defect segregation can modify grain-boundary resistance. The complex impedance of a polycrystalline ceramic may be idealized as
\begin{equation}
Z^*(\omega)=Z_g^*(\omega)+Z_{gb}^*(\omega)+Z_e^*(\omega),
\end{equation}
where the subscripts denote grain, grain-boundary, and electrode contributions \cite{r20}.

Microstructure records the full processing history of a ceramic. Grain growth is often described phenomenologically as
\begin{equation}
D^n-D_0^n=Kt,
\end{equation}
with
\begin{equation}
K=K_0\exp\left(-\frac{Q}{RT}\right),
\end{equation}
where $D$ is grain size, $n$ is a growth exponent, and $Q$ is an apparent activation energy. Porosity is another dominant variable. Mechanical strength often decreases approximately as
\begin{equation}
\sigma=\sigma_0e^{-bP},
\end{equation}
where $P$ is porosity fraction. Yet in biomedical scaffolds, interconnected porosity can improve fluid penetration, vascularization, cell migration, and tissue ingrowth, creating a fundamental porosity--strength trade-off \cite{r12,r19}.

\subsection{Structure--Property Relationships}
Ceramic properties may be expressed conceptually as
\begin{equation}
\mathcal{P}=f(C,S,D,M),
\end{equation}
where $C$ represents composition, $S$ crystal structure, $D$ defect state, and $M$ microstructure. Mechanical response is governed by strong bonding, limited dislocation mobility, and flaw sensitivity. Thermal behavior additionally depends on thermal conductivity, expansion, phase stability, and thermal-shock resistance. Thermal stress can be approximated as
\begin{equation}
\sigma_{th}\approx E\alpha\Delta T.
\end{equation}

Electrical behavior ranges from insulating to ionic-conducting, dielectric, ferroelectric, piezoelectric, and semiconducting regimes \cite{r17,r20,r39}. Complex permittivity is represented as
\begin{equation}
\varepsilon^*(\omega)=\varepsilon'(\omega)-j\varepsilon''(\omega),
\end{equation}
with
\begin{equation}
\tan\delta=\frac{\varepsilon''}{\varepsilon'}.
\end{equation}
Polarization can arise from electronic, ionic, dipolar, and interfacial mechanisms, with relative contributions depending on frequency, temperature, defects, and microstructure \cite{r17,r20,r39}.

Chemical stability is similarly environment dependent. Alumina is highly stable under many physiological conditions, whereas calcium phosphates and bioactive glasses are intentionally designed to dissolve or react. Degradation may be described conceptually as
\begin{equation}
r_d=f(A_s,pH,T,\mathrm{composition},\mathrm{crystallinity}),
\end{equation}
where $A_s$ is available surface area. For degradable implants, the design target is not maximum stability but approximate matching between material resorption and tissue regeneration.

\section{Processing and Manufacturing of Ceramic Materials}
\subsection{Powder Processing and Green-Body Formation}
Most engineering ceramics are manufactured through powder preparation, forming, drying or binder removal, and thermal consolidation. A simplified sequence is
$$
\text{Powder} \rightarrow \text{Conditioning}\rightarrow\text{Forming}\rightarrow\\
$$
$$
\text{Green\ Body}\rightarrow\text{Debinding}\
\rightarrow \text{Sintering}.
$$
Particle size distribution, morphology, surface chemistry, agglomeration, binder concentration, dispersants, and forming pressure influence green density and spatial homogeneity. Relative green density may be defined as
\begin{equation}
\rho_{g,r}=\frac{\rho_g}{\rho_{th}}.
\end{equation}
High average density is advantageous only when density gradients and agglomerates are simultaneously minimized \cite{r4,r6}.

\subsection{Sintering and Densification}
Sintering reduces surface and interfacial free energy through thermally activated mass transport. Neck growth may occur through surface diffusion, grain-boundary diffusion, lattice diffusion, vapor transport, or solution--precipitation. Not all mechanisms produce equivalent densification. Surface diffusion can enlarge necks without strongly reducing particle-center separation, whereas grain-boundary and lattice diffusion can contribute directly to shrinkage \cite{r6}.

The diffusion coefficient follows an Arrhenius form
\begin{equation}
D=D_0\exp\left(-\frac{Q_D}{RT}\right).
\end{equation}
Sintering represents a kinetic competition between densification and grain growth. Excessive grain coarsening may isolate pores and alter phase stability or dielectric response, so thermal schedules should be designed around the target combination of density, grain size, and phase composition rather than maximum exposure alone \cite{r4,r6}.

\subsection{Sol--Gel Processing}
Sol--gel processing provides molecular-scale compositional control and enables low-temperature synthesis, tailored porosity, and formation of coatings, glasses, hybrids, and nanostructured ceramics \cite{r7,r21}. A generic metal alkoxide may undergo hydrolysis and condensation:
\begin{align}
\mathrm{M-OR+H_2O}&\rightarrow\mathrm{M-OH+ROH},\\
\mathrm{M-OH+HO-M}&\rightarrow\mathrm{M-O-M+H_2O}.
\end{align}
The resulting sequence is
\begin{equation}
\mathrm{precursor}\rightarrow\mathrm{sol}\rightarrow\mathrm{gel}\rightarrow\mathrm{dry\ gel}\rightarrow\mathrm{ceramic}.
\end{equation}
Drying is critical because capillary pressure
\begin{equation}
P_c=\frac{2\gamma\cos\theta}{r}
\end{equation}
increases as pore radius decreases, explaining the tendency of fine-pore gels to crack during conventional drying. Supercritical drying can avoid liquid--vapor menisci and preserve highly porous aerogel structures \cite{r7}.

\subsection{Additive Manufacturing}
Ceramic additive manufacturing allows complex geometries to be generated directly from digital models but usually produces a green body that still requires debinding and sintering \cite{r8,r9,r40}. A generalized route is
$$
\text{CAD} \rightarrow \text{Feedstock} \rightarrow \text{Printing} \rightarrow 
$$
$$
\text{Green Part} \rightarrow \text{Debinding} \rightarrow \text{Sintering}
$$
Vat photopolymerization offers high resolution but requires a compromise between ceramic solids loading and slurry viscosity. Direct ink writing requires shear-thinning inks capable of flowing under extrusion while recovering sufficient yield stress after deposition. Binder-based and filament-based routes introduce additional concerns related to interlayer bonding, residual organics, voids, and anisotropic shrinkage \cite{r8,r22}.

Thermal debinding is particularly critical. If polymer decomposition exceeds volatile transport capacity, internal pressure can generate blistering, cracking, delamination, or distortion. A practical process condition is
\begin{equation}
r_{\mathrm{decomposition}}\lesssim r_{\mathrm{volatile\ transport}}.
\end{equation}
Recent reviews identify defect formation and long debinding times as central barriers in stereolithography-based ceramic manufacturing \cite{r9}.

Linear shrinkage after sintering may be quantified as
\begin{equation}
S_L=\frac{L_0-L_f}{L_0}\times100\%.
\end{equation}
Because shrinkage can be anisotropic, $S_x\neq S_y\neq S_z$, dimensional compensation may be required at the CAD stage. Multi-material printing further introduces mismatch in sintering shrinkage and thermal expansion, creating substantial interfacial-stress challenges \cite{r10}.

\subsection{Processing-Induced Defects}
Processing defects include residual pores, agglomerates, microcracks, delamination, incomplete interlayer bonding, binder-rich regions, residual carbon, phase heterogeneity, grain-size gradients, warping, and residual stresses. Final performance can be represented as
\begin{equation}
P=f(\rho,G,P_v,D_f,\sigma_r,\phi),
\end{equation}
where $\rho$ is density, $G$ grain size, $P_v$ pore volume fraction, $D_f$ defect population, $\sigma_r$ residual stress, and $\phi$ phase composition. Two components with similar density may display substantially different reliability if their flaw populations or phase states differ.

\section{Biomedical Ceramics and Biological Performance}
\subsection{Classification of Bioceramics}
Bioceramics may be broadly classified as bioinert, bioactive, or bioresorbable, although these categories are not absolute because biological response depends on surface condition, porosity, dopants, microstructure, and implantation environment \cite{r11,r41}. Their performance follows an interfacial sequence:
$$
\text{Bulk Composition} \rightarrow \text{Surface Chemistry} \rightarrow  
$$
$$
\text{Fluid Interaction} \rightarrow \text{Protein Adsorption} \rightarrow
$$
$$
\text{Cell Response} \rightarrow \text{Tissue Integration}
$$
Successful design requires simultaneous control of biological interaction and mechanical function.

\subsection{Alumina and Zirconia}
High-purity $\alpha$-Al$_2$O$_3$ is used because of hardness, wear resistance, chemical stability, and low physiological reactivity. Its performance depends strongly on purity, grain size, residual porosity, and flaw population. A major limitation is relatively low fracture toughness compared with transformation-toughened zirconia \cite{r11,r18}.

Zirconia combines high strength, fracture resistance, wear resistance, and favorable dental aesthetics. Stabilization of tetragonal zirconia with Y$_2$O$_3$ or other oxides enables transformation toughening:
$$
\text{crack stress} \rightarrow t \rightarrow m \rightarrow 
$$
$$
\text{local expansion} \rightarrow \text{compressive shielding}
$$
However, metastable tetragonal grains may progressively transform under humid conditions, leading to low-temperature degradation. Thus, metastability is simultaneously the source of toughening and a long-term stability concern \cite{r14,r15,r16}.

\subsection{Calcium Phosphate Ceramics}
Hydroxyapatite has nominal composition
\begin{equation}
\mathrm{Ca_{10}(PO_4)_6(OH)_2}
\end{equation}
and ideal Ca/P atomic ratio of approximately 1.67. Its chemical relationship with bone mineral gives it strong osteoconductive potential, but its brittleness and limited toughness restrict monolithic load-bearing use \cite{r12,r19,r42}. Tricalcium phosphate, Ca$_3$(PO$_4$)$_2$, is generally more soluble and is useful when greater resorbability is required. The design requirement is approximate matching
\begin{equation}
r_{\mathrm{resorption}}\approx r_{\mathrm{tissue\ formation}}.
\end{equation}
Biphasic calcium phosphate systems combine hydroxyapatite and $\beta$-TCP to tune degradation and biological response \cite{r12,r19,r42}.

\subsection{Bioactive Glasses}
Bioactive glasses are designed to undergo controlled surface reactions in physiological fluids. Classical 45S5 Bioglass contains approximately 45 wt\% SiO$_2$, 24.5 wt\% Na$_2$O, 24.5 wt\% CaO, and 6 wt\% P$_2$O$_5$ \cite{r23,r24}. A simplified reaction sequence is
$$
\text{ion exchange} \rightarrow \text{network hydrolysis} \rightarrow
$$
$$
\text{silica layer} \rightarrow \text{Ca-P enrichment} \rightarrow \text{apatite-like layer}
$$
Silicate hydrolysis may be written schematically as
\begin{equation}
\mathrm{Si-O-Si+H_2O}\rightarrow\mathrm{Si-OH+HO-Si}.
\end{equation}
Modern bioactive-glass research extends beyond bone bonding toward controlled therapeutic-ion release and angiogenic, antibacterial, and immunomodulatory functions \cite{r13,r24,r25}.

\subsection{Porosity, Degradation, and Tissue Integration}
Porosity increases surface area, fluid penetration, cell-accessible volume, and often degradation rate, but simultaneously lowers modulus and strength. The central scaffold trade-off is
\begin{equation}
\mathrm{Biological\ Accessibility}\leftrightarrow\mathrm{Mechanical\ Integrity}.
\end{equation}
Pore connectivity can be as important as total porosity because isolated pores contribute little to vascularization or tissue ingrowth. Additive manufacturing is particularly attractive because it allows pore architecture to be designed rather than inherited stochastically from processing \cite{r12,r22}.

\subsection{Dental and Orthopedic Applications}
Dental ceramics must balance strength, wear resistance, chemical stability, biocompatibility, and optical compatibility. Zirconia illustrates a characteristic compromise between translucency and transformation-toughening capacity as phase composition shifts toward higher cubic fractions \cite{r14}. Orthopedic ceramics follow two broad paradigms: permanent structural ceramics, in which long-term fracture and wear resistance dominate, and regenerative ceramics, in which controlled degradation and biological signaling are required. Alumina and zirconia serve different functions from calcium phosphates and bioactive glasses even when all are classified as bioceramics \cite{r11,r13,r41}.

\section{Advanced and Emerging Ceramic Systems}
\subsection{Nanoceramics}
Reducing ceramic dimensions into the nanoscale increases surface-to-volume ratio approximately as
\begin{equation}
\frac{A}{V}\propto\frac{1}{d},
\end{equation}
thereby increasing the influence of surfaces, interfaces, and grain boundaries. This can modify diffusion, sintering kinetics, phase stability, dissolution, and biological interactions. Nanostructuring is particularly relevant in calcium-phosphate systems because natural bone mineral itself contains nanoscale features. Yet nanoparticle agglomeration and rapid grain growth create a central densification--grain-growth conflict during consolidation \cite{r12,r26}.

\subsection{Smart and Responsive Ceramics}
Piezoelectric ceramics convert mechanical stress to electrical displacement:
\begin{equation}
D_i=d_{ijk}\sigma_{jk},
\end{equation}
and electric field to strain:
\begin{equation}
S_{jk}=d_{ijk}E_i.
\end{equation}
These relationships underlie sensors, actuators, ultrasonic transducers, and energy harvesters. Ferroelectric ceramics add switchable spontaneous polarization, while zirconia demonstrates a mechanically responsive phase transformation. Smart behavior can be interpreted broadly as controlled coupling between external stimuli and structural, electrical, optical, or mechanical response \cite{r27}.

\subsection{Ceramic Matrix and Hybrid Composites}
Composite design introduces crack deflection, crack bridging, transformation toughening, interfacial debonding, and fiber pull-out as mechanisms that can increase damage tolerance. Crack shielding may be represented as
\begin{equation}
K_{tip}=K_{applied}-K_{shielding}.
\end{equation}
Zirconia-toughened alumina is a representative biomedical composite combining alumina wear resistance with zirconia transformation toughening \cite{r18}.

Ceramic--polymer hybrids are particularly attractive in biomedical engineering because ceramics provide stiffness, bioactivity, and mineral chemistry while polymers contribute toughness, flexibility, processability, and controlled degradation. Their success depends strongly on interfacial adhesion and load transfer.

\subsection{Defect and Interface Engineering}
Modern functional ceramics increasingly treat defects as design variables rather than imperfections to be eliminated. Dopants, oxygen vacancies, grain boundaries, interfaces, and controlled porosity can modify ion transport, surface charge, dissolution, and biological activity. This represents a transition from
\begin{equation}
\mathrm{defect\ minimization}\rightarrow\mathrm{defect\ control}.
\end{equation}
Recent bioceramic literature specifically emphasizes the combined roles of defect structures and interfaces in osteointegration, cell adhesion, ion release, and antibacterial functionality \cite{r28}.

\subsection{Multimaterial and Architected Ceramics}
Multimaterial additive manufacturing enables composition gradients and spatially programmed properties but requires compatibility in rheology, debinding, sintering kinetics, and thermal expansion. Mismatch may be described by
\begin{equation}
\Delta S=S_1-S_2, \qquad \Delta\alpha=\alpha_1-\alpha_2,
\end{equation}
which can generate interfacial stresses during co-sintering and cooling \cite{r10}. Architected ceramics additionally exploit geometry as a design variable. Cellular solids often obey scaling relationships of the form
\begin{equation}
\frac{E^*}{E_s}=C\left(\frac{\rho^*}{\rho_s}\right)^n,
\end{equation}
allowing effective stiffness, permeability, and mass to be modified without changing composition \cite{r29}.

\section{Sustainability and Circular Manufacturing}
\subsection{Environmental Burden and Circularity}
Conventional ceramic production combines mineral extraction, transportation, forming, drying, and high-temperature firing. A simplified thermal requirement is
\begin{equation}
Q=mC_p\Delta T+Q_{reaction}+Q_{loss}.
\end{equation}
Sustainability depends on raw-material conservation, reduction of firing temperature and time, furnace efficiency, low-carbon energy, waste recovery, and product lifetime \cite{r30,r31}.

A circular ceramic model seeks to transform waste streams into secondary raw materials:
\begin{equation}
\begin{aligned}
\mathrm{Raw\ Material} &\rightarrow \mathrm{Product}\rightarrow\mathrm{Recovery}\\
&\rightarrow \mathrm{Secondary\ Raw\ Material}.
\end{aligned}
\end{equation}
Potential sources include fly ash, red mud, blast-furnace slag, waste glass, ceramic scrap, metallurgical residues, industrial sludge, and agricultural ashes \cite{r30,r32}.

\subsection{Waste-Derived Feedstocks}
Waste glass can participate in viscous sintering and may reduce required firing temperatures, whereas red mud can provide Fe$_2$O$_3$, Al$_2$O$_3$, SiO$_2$, TiO$_2$, and alkali-containing feedstock. Recycled ceramic and alumina powders may also reduce virgin-material demand, but compositional variability, trace contaminants, particle-size differences, and altered sintering kinetics must be carefully controlled \cite{r30,r32,r33}.

Recycled content should not be used as a standalone sustainability metric. A more meaningful objective is
\begin{equation}
\begin{aligned}
&\mathrm{minimum\ environmental\ impact}\\
&\mathrm{per\ unit\ functional\ lifetime}.
\end{aligned}
\end{equation}
This is particularly important for biomedical ceramics, where purity, traceability, sterilizability, and reliability constrain the use of heterogeneous secondary feedstocks.

\section{Current Challenges and Limitations}
\subsection{Brittle Fracture and Statistical Reliability}
Ceramic strength is statistical because failure is governed by flaw populations. Weibull analysis is commonly used:
\begin{equation}
P_f=1-\exp\left[-\left(\frac{\sigma}{\sigma_0}\right)^m\right],
\end{equation}
where $P_f$ is failure probability, $\sigma_0$ characteristic strength, and $m$ the Weibull modulus. Advanced ceramic engineering must optimize not only mean strength but also strength distribution, flaw populations, and long-term reliability \cite{r4,r5}.

\subsection{Porosity--Strength and Degradation--Regeneration Trade-Offs}
For structural ceramics, increasing porosity generally lowers strength and modulus. For regenerative scaffolds, however, porosity is required for cell migration, nutrient transport, vascularization, and tissue ingrowth. Similarly, resorbable ceramics require degradation but must maintain sufficient mechanical function until tissue replacement occurs. Thus,
\begin{equation}
\mathrm{Porosity}\leftrightarrow\mathrm{Mechanical\ Strength},
\end{equation}
and
\begin{equation}
\mathrm{Bioactivity}\leftrightarrow\mathrm{Structural\ Persistence}
\end{equation}
remain central design conflicts.

\subsection{Processing Defects and Scale-Up}
Additive manufacturing introduces additional defects associated with curing, interlayer bonding, debinding, anisotropic shrinkage, warping, and thermal gradients. Scaling to larger components increases transport distances and makes thermal and shrinkage gradients more difficult to control. Geometric freedom remains coupled to process complexity \cite{r8,r9}.

\subsection{Long-Term Biomedical Stability}
Zirconia illustrates the need for long-term validation. Humid environments can promote low-temperature degradation and surface transformation, while accelerated laboratory tests may not always reproduce long-term clinical kinetics. Five-year clinical evidence has confirmed detectable aging behavior in 3Y-TZP dental prostheses \cite{r15,r16}. Likewise, in vitro apatite formation or cytocompatibility cannot by itself demonstrate long-term in vivo integration.

\subsection{Standardization and Translation}
Direct comparison among studies is complicated by differences in specimen geometry, surface finish, porosity, grain size, loading rate, fluid composition, and degradation conditions. Standardized reporting of processing, microstructure, test conditions, and biological environment is necessary both for reproducibility and for reliable data-driven modeling. Clinical translation additionally requires progression from materials characterization through in vitro, in vivo, and long-term clinical validation.

\section{Future Directions}
\subsection{Integrated Modeling and Data-Driven Design}
Future advances in ceramic science will increasingly require predictive frameworks capable of connecting processing parameters to phase evolution, microstructure, defects, and final performance. Data-driven models can be written conceptually as
\begin{equation}
\mathbf{y}=F(\mathbf{x}),
\end{equation}
where $\mathbf{x}$ contains processing and compositional variables and $\mathbf{y}$ contains predicted properties. Machine learning is increasingly used for microstructure-informed materials modeling, inverse design, and property prediction, but physically meaningful extrapolation requires integration with mechanistic constraints \cite{r34}.

This integration can be summarized conceptually as
\begin{equation}
\mathrm{Experimental\ Data}+\mathrm{Physics}+\mathrm{Machine\ Learning}.
\end{equation}
Digital-twin concepts may further integrate sensor data, process models, and real-time control for sintering or additive manufacturing.

\subsection{In Situ Characterization}
In situ diffraction, tomography, Raman spectroscopy, electron microscopy, and impedance measurements can reveal dynamic phase transformations, shrinkage, pore evolution, transport, and crack initiation. Rather than comparing only initial and final states, future work should increasingly resolve
\begin{equation}
\phi_i=\phi_i(T,t), \qquad P=P(T,t),
\end{equation}
where $\phi_i$ is phase fraction and $P$ porosity. Such measurements are especially relevant to transformation-toughened zirconia and thermally processed additive-manufactured ceramics.

\subsection{Personalized and 4D-Printed Bioceramics}
Patient-specific manufacturing can integrate medical imaging, CAD, additive manufacturing, and spatially designed porosity or stiffness. The next step is to move from patient-specific geometry toward patient-specific function. Four-dimensional printing introduces time-dependent, stimulus-responsive behavior into 3D structures and is being explored for adaptive bone implants, although mechanical stability and reproducibility remain substantial barriers \cite{r35}.

\subsection{Therapeutic-Ion and Immunomodulatory Ceramics}
Advanced bioceramics are increasingly designed to release ions such as Mg$^{2+}$, Sr$^{2+}$, Zn$^{2+}$, Cu$^{2+}$, or Si-containing species to influence osteogenesis, angiogenesis, antibacterial activity, or immune response. The design target is an appropriate concentration window rather than maximum release. Recent work emphasizes the importance of immune regulation, particularly macrophage-mediated responses, in successful bone regeneration \cite{r36,r37,r38}.

This suggests an evolution from passive scaffolds toward materials that combine mechanical support with biochemical regulation. However, immune responses are time dependent and context specific; future systems may require spatial and temporal control of degradation and ion release.

\subsection{Functionally Graded and Multimaterial Implants}
Natural tissues exhibit gradients in density, mineralization, porosity, and stiffness. Functionally graded ceramics can vary composition or architecture according to position:
\begin{equation}
C=C(x),\qquad P=P(x),\qquad E=E(x).
\end{equation}
Such structures may combine dense load-bearing zones with porous bioactive interfaces, provided co-sintering, shrinkage, and thermal-expansion compatibility can be controlled \cite{r10}.

\subsection{Low-Energy and Circular Processing}
Future ceramic studies should increasingly report performance and sustainability together. A comprehensive dataset might include
\begin{equation}
\rho,\ G,\ K_{IC},\ \sigma_f,\ E_s,\ CO_{2e},
\end{equation}
where $E_s$ is specific process energy and $CO_{2e}$ carbon-equivalent emissions. Rapid sintering, pressure-assisted densification, improved kiln efficiency, renewable-electric heating, near-net-shape manufacturing, and carefully selected secondary feedstocks can reduce environmental impact while preserving required microstructures \cite{r30,r31,r32}.

\section{Conclusions}
Ceramic materials have evolved from traditional structural materials into highly engineered systems whose performance is governed by the interaction among composition, atomic bonding, crystal structure, defects, interfaces, microstructure, processing history, and architecture. The central conclusion of this review is that ceramic behavior cannot be interpreted adequately from nominal composition or material class alone. Instead, performance emerges from a hierarchical relationship:
$$
\text{Composition} \rightarrow \text{Bonding}\rightarrow\text{Crystal Structure}\rightarrow\\
$$
$$
\text{Defects}\rightarrow\text{Microstructure}\rightarrow\\
\text{Properties}\rightarrow\text{Performance}
$$
Processing is a design variable rather than merely a fabrication step. Conventional powder routes, sintering, sol--gel processing, and additive manufacturing generate distinct phase states, defect populations, and microstructures. Additive manufacturing expands geometric freedom, but feedstock rheology, debinding, shrinkage, interlayer quality, and residual porosity remain central barriers to reliable structural and biomedical components.

Biomedical ceramics provide a particularly clear demonstration of coupled optimization. Alumina and zirconia emphasize long-term structural reliability, whereas hydroxyapatite, tricalcium phosphate, and bioactive glasses rely on controlled chemical interaction with the physiological environment. Zirconia illustrates the duality of metastability: the tetragonal phase enables transformation toughening, yet the same metastability contributes to aging susceptibility. Calcium phosphates and bioactive glasses illustrate a different design condition in which degradation is not necessarily failure but must be coordinated with tissue regeneration.

Porosity similarly changes meaning with application. In structural ceramics it is generally a flaw population; in regenerative scaffolds it becomes a functional architecture. Consequently, ceramic design increasingly requires optimization across competing variables rather than maximization of a single property. This principle extends to sustainable manufacturing, where recycled content or lower firing temperature is beneficial only when functional lifetime and reproducibility are preserved.

The field is now transitioning from monolithic materials toward nanoceramics, smart ceramics, ceramic composites, defect-engineered systems, multimaterial architectures, and immunomodulatory bioceramics. Vacancies, grain boundaries, interfaces, dopants, and architecture are increasingly treated as controllable variables. Future progress will depend on integrating advanced processing, in situ characterization, predictive modeling, and data-driven design while maintaining physically grounded interpretation.

Overall, next-generation ceramics should be regarded as hierarchically engineered systems in which atomic chemistry, defects, interfaces, microstructure, and architecture are coordinated across length scales to produce application-specific performance. This integrated perspective provides a foundation for designing ceramic materials that are not only stronger or more bioactive, but also more predictable, multifunctional, manufacturable, sustainable, and clinically relevant.


\begin{thebibliography}{00}

\bibitem{r1}
M. W. Barsoum, \emph{Fundamentals of Ceramics}. Boca Raton, FL, USA: CRC Press, 2019.
\bibitem{r2}
C. B. Carter and M. G. Norton, \emph{Ceramic Materials: Science and Engineering}, 2nd ed. New York, NY, USA: Springer, 2013.
\bibitem{r3}
W. D. Kingery, H. K. Bowen, and D. R. Uhlmann, \emph{Introduction to Ceramics}, 2nd ed. New York, NY, USA: Wiley, 1976.
\bibitem{r4}
M. N. Rahaman, \emph{Ceramic Processing and Sintering}, 2nd ed. Boca Raton, FL, USA: CRC Press, 2003.
\bibitem{r5}
B. Lawn, \emph{Fracture of Brittle Solids}, 2nd ed. Cambridge, U.K.: Cambridge University Press, 1993.
\bibitem{r6}
R. M. German, \emph{Sintering Theory and Practice}. New York, NY, USA: Wiley, 1996.
\bibitem{r7}
P. Innocenzi, ``Sol-gel processing for advanced ceramics, a perspective,'' \emph{Open Ceramics}, vol. 16, Art. no. 100477, 2023, doi: 10.1016/j.oceram.2023.100477.
\bibitem{r8}
M. Dadkhah, J.-M. Tulliani, A. Saboori, and L. Iuliano, ``Additive manufacturing of ceramics: Advances, challenges, and outlook,'' \emph{J. Eur. Ceram. Soc.}, vol. 43, no. 15, pp. 6635--6664, 2023, doi: 10.1016/j.jeurceramsoc.2023.07.033.
\bibitem{r9}
S. Zhou, G. Liu, C. Wang, Y. Zhang, C. Yan, and Y. Shi, ``Thermal debinding for stereolithography additive manufacturing of advanced ceramic parts: A comprehensive review,'' \emph{Mater. Des.}, vol. 238, Art. no. 112632, 2024, doi: 10.1016/j.matdes.2024.112632.
\bibitem{r10}
H. Sun, B. Zou, X. Wang, W. Chen, G. Zhang, T. Quan, and C. Huang, ``Advancements in multi-material additive manufacturing of advanced ceramics: A review of strategies, techniques and equipment,'' \emph{Mater. Chem. Phys.}, vol. 319, Art. no. 129337, 2024, doi: 10.1016/j.matchemphys.2024.129337.
\bibitem{r11}
L. Vaiani, A. Boccaccio, A. E. Uva, G. Palumbo, A. Piccininni, P. Guglielmi, S. Cantore, L. Santacroce, I. A. Charitos, and A. Ballini, ``Ceramic materials for biomedical applications: An overview on properties and fabrication processes,'' \emph{J. Funct. Biomater.}, vol. 14, no. 3, Art. no. 146, 2023, doi: 10.3390/jfb14030146.
\bibitem{r12}
F. Fendi, B. Abdullah, S. Suryani, A. N. Usman, and D. Tahir, ``Development and application of hydroxyapatite-based scaffolds for bone tissue regeneration: A systematic literature review,'' \emph{Bone}, vol. 183, Art. no. 117075, 2024, doi: 10.1016/j.bone.2024.117075.
\bibitem{r13}
D. M. B. Nogueira, M. P. de O. Rosso, D. V. Buchaim, M. S. R. Zangrando, and R. L. Buchaim, ``Update on the use of 45S5 bioactive glass in the treatment of bone defects in regenerative medicine,'' \emph{World J. Orthop.}, vol. 15, no. 3, pp. 204--214, 2024, doi: 10.5312/wjo.v15.i3.204.
\bibitem{r14}
C. H. Lim, S. Vardhaman, N. Reddy, and Y. Zhang, ``Composition, processing, and properties of biphasic zirconia bioceramics: Relationship to competing strength and optical properties,'' \emph{Ceram. Int.}, vol. 48, no. 12, pp. 17095--17103, 2022, doi: 10.1016/j.ceramint.2022.02.265.
\bibitem{r15}
V. Koenig, T. Douillard, J. Chevalier, F. Amiard, M. Lamy de la Chapelle, S. Le Goff, A. Vanheusden, N. Dardenne, C. Wulfman, and A. Mainjot, ``Intraoral low-temperature degradation of monolithic zirconia dental prostheses: 5-year results of a prospective clinical study with ex vivo monitoring,'' \emph{Dent. Mater.}, vol. 40, no. 2, pp. 198--209, 2024, doi: 10.1016/j.dental.2023.11.008.
\bibitem{r16}
J. Chevalier, ``What future for zirconia as a biomaterial?,'' \emph{Biomaterials}, vol. 27, no. 4, pp. 535--543, 2006, doi: 10.1016/j.biomaterials.2005.07.034.
\bibitem{r17}
M. Zhang, R. Wei, J. Zeng, and Y.-H. Lin, ``Probing atomic-scale structure of dielectric ceramics with scanning transmission electron microscopy,'' \emph{J. Adv. Dielectr.}, vol. 14, no. 3, 2024, doi: 10.1142/S2010135X23430014.
\bibitem{r18}
C. Piconi and G. Maccauro, ``Zirconia as a ceramic biomaterial,'' \emph{Biomaterials}, vol. 20, no. 1, pp. 1--25, 1999.
\bibitem{r19}
S. V. Dorozhkin, ``Bioceramics of calcium orthophosphates,'' \emph{Biomaterials}, vol. 31, no. 7, pp. 1465--1485, 2010, doi: 10.1016/j.biomaterials.2009.11.050.
\bibitem{r20}
T. Defferriere, D. Klotz, J. C. Gonzalez-Rosillo, J. L. M. Rupp, and H. L. Tuller, ``Photo-enhanced ionic conductivity across grain boundaries in polycrystalline ceramics,'' \emph{Nat. Mater.}, vol. 21, no. 4, pp. 438--444, 2022, doi: 10.1038/s41563-021-01181-2.
\bibitem{r39}
D. Li, D. Jia, Z. Yang, and Y. Zhou, ``Principles, design, structure and properties of ceramics for microwave absorption or transmission at high temperatures,'' \emph{Int. Mater. Rev.}, vol. 67, no. 3, pp. 266--297, 2022, doi: 10.1080/09506608.2021.1941716.
\bibitem{r21}
K. Kajihara, K. Kanamori, and A. Shimojima, ``Current status of sol--gel processing of glasses, ceramics, and organic--inorganic hybrids: A brief review,'' \emph{J. Ceram. Soc. Jpn.}, vol. 130, no. 8, pp. 575--583, 2022, doi: 10.2109/jcersj2.22078.
\bibitem{r40}
J. S. Pelz, N. Ku, M. A. Meyers, and L. R. Vargas-Gonzalez, ``Additive manufacturing of structural ceramics: A historical perspective,'' \emph{J. Mater. Res. Technol.}, vol. 15, pp. 670--695, 2021, doi: 10.1016/j.jmrt.2021.07.155.
\bibitem{r22}
E. Romanczuk-Ruszuk, B. Sztorch, D. Pakula, E. Gabriel, K. Nowak, and R. E. Przekop, ``3D printing ceramics---Materials for direct extrusion process,'' \emph{Ceramics}, vol. 6, no. 1, pp. 364--385, 2023, doi: 10.3390/ceramics6010022.
\bibitem{r41}
E. A. Ibarra, R. Adames, J. Castro, S. Guevara, L. E. Petrocelli, D. Reginensi, J. Molino, and M. Vasquez, ``Avances en biocer\'amicas para la regeneraci\'on \'osea: De materiales bioinertes a compuestos bioactivos,'' \emph{Eur. Sci. J.}, vol. 20, no. 15, p. 1, 2024, doi: 10.19044/esj.2024.v20n15p1.
\bibitem{r42}
S. V. Dorozhkin, ``Calcium orthophosphate-based bioceramics,'' \emph{Materials}, vol. 6, no. 9, pp. 3840--3942, 2013, doi: 10.3390/ma6093840.
\bibitem{r23}
L. L. Hench, ``The story of Bioglass,'' \emph{J. Mater. Sci.: Mater. Med.}, vol. 17, pp. 967--978, 2006, doi: 10.1007/s10856-006-0432-z.
\bibitem{r24}
M. N. Rahaman, D. E. Day, B. S. Bal, Q. Fu, S. B. Jung, L. F. Bonewald, and A. P. Tomsia, ``Bioactive glass in tissue engineering,'' \emph{Acta Biomater.}, vol. 7, no. 6, pp. 2355--2373, 2011, doi: 10.1016/j.actbio.2011.03.016.
\bibitem{r25}
T. Kokubo and H. Takadama, ``How useful is SBF in predicting in vivo bone bioactivity?,'' \emph{Biomaterials}, vol. 27, no. 15, pp. 2907--2915, 2006, doi: 10.1016/j.biomaterials.2006.01.017.
\bibitem{r26}
M. Saha and M. Mallik, ``3D printing of nanoceramics: Present status and future perspectives,'' preprint, 2023, doi: 10.31224/2884.
\bibitem{r27}
M. Mhadhbi and A. Sohani, ``Introductory chapter: Smart and advanced ceramics and applications,'' in \emph{Smart and Advanced Ceramics and Applications}. London, U.K.: IntechOpen, 2022, doi: 10.5772/intechopen.105977.
\bibitem{r28}
N. Anwar and M. Ahmed, ``Interface and defect engineering in bioceramic materials for biomedical applications: A review,'' \emph{Ceram. Int.}, in press, 2026, doi: 10.1016/j.ceramint.2026.08.069.
\bibitem{r29}
L. J. Gibson and M. F. Ashby, \emph{Cellular Solids: Structure and Properties}, 2nd ed. Cambridge, U.K.: Cambridge University Press, 1997.
\bibitem{r30}
Z. Li, W. Li, J. You, J. Huang, R. Gan, J. Guo, and X. Zhang, ``Critical secondary resource for porous ceramics: A review on recycling of inorganic solid wastes,'' \emph{J. Eur. Ceram. Soc.}, vol. 44, no. 15, Art. no. 116781, 2024, doi: 10.1016/j.jeurceramsoc.2024.116781.
\bibitem{r31}
S. S. Hossain and P. K. Roy, ``Sustainable ceramics derived from solid wastes: A review,'' \emph{J. Asian Ceram. Soc.}, vol. 8, no. 4, pp. 984--1009, 2020, doi: 10.1080/21870764.2020.1815348.
\bibitem{r32}
I. Zmak and L. \'{C}urkovi\'{c}, ``Improving sustainability of technical ceramics production: Synergistic approach,'' \emph{ENTRENOVA}, vol. 7, no. 1, pp. 376--384, 2021, doi: 10.54820/jojw7514.
\bibitem{r33}
U. Parmar, S. Ahmed, R. Vayas, C. Lavanya, M. Manjunatha, N. Singh, and H. Kumar, ``Towards sustainable ceramic forming: Techniques, materials, and applications in evolving paradigms,'' \emph{E3S Web Conf.}, vol. 552, Art. no. 01099, 2024, doi: 10.1051/e3sconf/202455201099.
\bibitem{r34}
X.-L. Peng, M. Fathidoost, B. Lin, Y. Yang, and B.-X. Xu, ``What can machine learning help with microstructure-informed materials modeling and design?,'' \emph{MRS Bull.}, vol. 50, pp. 61--79, 2025, doi: 10.1557/s43577-024-00797-4.
\bibitem{r35}
F. Yang, W. Ding, J. Jia, C. Shuai, and P. Feng, ``4D printing for bone implant: Progress, advantages and challenges,'' \emph{Prog. Mater. Sci.}, vol. 157, Art. no. 101591, 2026, doi: 10.1016/j.pmatsci.2025.101591.
\bibitem{r36}
J. Li, Y. Qu, B. Chu, T. Wu, M. Pan, D. Mo, L. Li, Y. Ming, Y. Yang, M. Wang, X. He, and Z. Qian, ``Research progress on biomaterials with immunomodulatory effects in bone regeneration,'' \emph{Adv. Sci.}, vol. 12, no. 33, Art. no. e01209, 2025, doi: 10.1002/advs.202501209.
\bibitem{r37}
Y. Hu, Y. Xu, Q. Shen, S. Yu, H. Xiong, and C. Fan, ``Current state of bioceramic bone repair materials in immune regulation: A review,'' \emph{Front. Med.}, vol. 19, no. 6, pp. 911--932, 2025, doi: 10.1007/s11684-025-1179-6.
\bibitem{r38}
A. Sarasati, I. D. Ana, D. Yucel, H. Kenar, H. Wihadmadyatami, N. S. Ozdemir, R. Anggraeni, and V. Hasirci, ``Surface functionalized calcium phosphate bioceramics for immunomodulatory biomaterials,'' \emph{Dent. Mater. J.}, vol. 44, no. 6, pp. 599--617, 2025, doi: 10.4012/dmj.2025-100.

\end{thebibliography}
\end{document}